# A New Sphere-Packing Bound for Maximal Error Exponent for Multiple-Access Channels

Ali Nazari, Sandeep Pradhan and Achilleas Anastasopoulos
Electrical Engineering and Computer Science Dept.
University of Michigan, Ann Arbor, MI 48109-2122, USA
E-mail: {anazari,pradhanv,anastas}@umich.edu

November 19, 2018

## Abstract

In this work, a new lower bound for the maximal error probability of a two-user discrete memoryless (DM) multiple-access channel (MAC) is derived. This is the first bound of this type that explicitly imposes independence of the users' input distributions (conditioned on the time-sharing auxiliary variable) and thus results in a tighter sphere-packing exponent when compared to the tightest known exponent derived by Haroutunian.

## 1 introduction

An interesting problem in network information theory is to determine the minimum probability of error which can be achieved on a discrete memoryless (DM), multiple-access channel (MAC). More specifically, a two-user DM-MAC is defined by a stochastic matrix[1] $W : \mathcal{X} \times \mathcal{Y} \to \mathcal{Z}$, where the input alphabets, $\mathcal{X}, \mathcal{Y}$, and the output alphabet, $\mathcal{Z}$, are finite sets. The channel transition probability for sequences of length $n$ is given by

$$W^n(\mathbf{z}|\mathbf{x}, \mathbf{y}) \triangleq \prod_{i=1}^{n} W(z_i|x_i, y_i) \qquad (1)$$

where

$$\mathbf{x} \triangleq (x_1, ..., x_n) \in \mathcal{X}^n, \mathbf{y} \triangleq (y_1, ..., y_n) \in \mathcal{Y}^n$$

and

$$\mathbf{z} \triangleq (z_1, ..., z_n) \in \mathcal{Z}^n.$$

---

[1] We use the following notation throughout this work. Script capitals $\mathcal{U}, \mathcal{X}, \mathcal{Y}, \mathcal{Z}, \ldots$ denote finite, nonempty sets. To show the cardinality of a set $\mathcal{X}$, we use $|\mathcal{X}|$. We also use the letters $P, Q, \ldots$ for probability distributions on finite sets, and $U, X, Y, \ldots$ for random variables.

It is known [1], that for any $(R_X, R_Y)$ in the interior of a certain set $\mathcal{C}$, and for all sufficiently large $n$, there exists a multiuser code with an arbitrary small average probability of error. Conversely, for any $(R_X, R_Y)$ outside of $\mathcal{C}$, the average probability of error is bounded away from 0. The set $\mathcal{C}$, which is called *capacity region* for $W$, is the closure of the set of all rate pairs $(R_X, R_Y)$ satisfying [2]

$$0 \leq R_X \leq I(X \wedge Z|Y, Q) \tag{2a}$$
$$0 \leq R_Y \leq I(Y \wedge Z|X, Q) \tag{2b}$$
$$0 \leq R_X + R_Y \leq I(XY \wedge Z|Q), \tag{2c}$$

for all choices of joint distributions over the random variables $Q$, $X$, $Y$, $Z$ of the form $P(q)P(x|q)P(y|q)W(z|x,y)$ with $Q \in \mathcal{Q}$ and $|\mathcal{Q}| \leq 4$.

Haroutunian [3] derived a *lower* bound on the optimal average error probability for $W$. This result asserts that the average probability of error is bounded below by $\exp\{-nE_{sp}(R_X, R_Y, W)\}$, where

$$E_{sp}(R_X, R_Y, W) \triangleq \max_{P_{XY}} \min_{V_{Z|XY}} D(V_{Z|XY}||W|P_{XY}). \tag{3}$$

Here, the maximum is taken over all possible joint distributions over the random variables $X$, $Y$, and the minimum over all test channels $V_{Z|XY}$ which satisfy at least one of the following conditions

$$I_V(X \wedge Z|Y) \leq R_X \tag{4a}$$
$$I_V(Y \wedge Z|X) \leq R_Y \tag{4b}$$
$$I_V(XY \wedge Z) \leq R_X + R_Y, \tag{4c}$$

where $V \triangleq V_{Z|XY} \times P_{XY}$. This bound tends to be somewhat loose because it does not take into account the separation of the two encoders in the MAC.

In this paper, we derive a new lower bound that explicitly captures the separation of the encoders in the MAC and thus is tighter than the one provided by Haroutunian. However, this bound is only valid for the maximal and not the average error probability. Nevertheless, we believe that the techniques used in this derivation can be extended to provide lower bounds for the average error probability as well.

The paper is organized as follows. First, some preliminaries are introduced in section 2. Then in section 3, we state and prove the main result. The proof hinges upon a strong converse theorem which is also stated in the same Section and proved in the Appendix.

## 2 Preliminaries

For any alphabet $\mathcal{X}$, $\mathcal{P}(\mathcal{X})$ denotes the set of all probability distributions on $\mathcal{X}$. The *type* of a sequence $\mathbf{x} = (x_1, ..., x_n) \in \mathcal{X}^n$ is the distributions $P_\mathbf{x}$ on $\mathcal{X}$ defined by

$$P_\mathbf{x}(x) \triangleq \frac{1}{n}N(x|\mathbf{x}), \qquad x \in \mathcal{X}, \tag{5}$$

where $N(x|\mathbf{x})$ denotes the number of occurrences of $x$ in $\mathbf{x}$. Let $\mathcal{P}_n(\mathcal{X})$ denote the set of all types in $\mathcal{X}^n$, and define the set of all sequences in $\mathcal{X}^n$ of type $P$ as

$$T_P \triangleq \{\mathbf{x} \in \mathcal{X}^n : P_\mathbf{x} = P\}. \tag{6}$$

The joint type of a pair $(\mathbf{x}, \mathbf{y}) \in \mathcal{X}^n \times \mathcal{Y}^n$ is the probability distribution $P_{\mathbf{x},\mathbf{y}}$ on $\mathcal{X} \times \mathcal{Y}$ defined by

$$P_{\mathbf{x},\mathbf{y}}(x,y) \triangleq \frac{1}{n} N(x,y|\mathbf{x},\mathbf{y}), \qquad (x,y) \in \mathcal{X} \times \mathcal{Y}, \tag{7}$$

where $N(x,y|\mathbf{x},\mathbf{y})$ is the number of occurrences of $(x,y)$ in $(\mathbf{x},\mathbf{y})$. The relative entropy or *Kullback Leibler* distance between two probability distribution $P, Q \in \mathcal{P}(\mathcal{X})$ is defined as

$$D(P||Q) \triangleq \sum_{x \in \mathcal{X}} P(x) \log \frac{P(x)}{Q(x)}. \tag{8}$$

Let $\mathcal{W}(\mathcal{Y}|\mathcal{X})$ denote the set of all stochastic matrices with input alphabet $\mathcal{X}$ and output alphabet $\mathcal{Y}$. Then, given stochastic matrices $V, W \in \mathcal{W}(\mathcal{Y}|\mathcal{X})$, the conditional *I-divergence* is defined by

$$D(V||W|P) \triangleq \sum_{x \in \mathcal{X}} P(x) D(V(\cdot|x)||W(\cdot|x)). \tag{9}$$

An $(n, M, \lambda)$ code for $W : \mathcal{X} \to \mathcal{Z}$, is a system $\{(\mathbf{u}_i, D_i) : 1 \leq i \leq M\}$ with

- $\mathbf{u}_i \in \mathcal{X}^n$, $D_i \subset \mathcal{Z}^n$
- $D_i \cap D_{i'} = \varnothing$ for $i \neq i'$
- $W^n(D_i|\mathbf{u}_i) \geq 1 - \lambda$, for $1 \leq i \leq M$.

An $(n, M, N)$ multi-user code is a set $\{(\mathbf{u}_i, \mathbf{v}_j, D_{ij}) : 1 \leq i \leq M, 1 \leq j \leq N\}$ with

- $\mathbf{u}_i \in \mathcal{X}^n$, $\mathbf{v}_j \in \mathcal{Y}^n$, $D_{ij} \subset \mathcal{Z}^n$
- $D_{ij} \cap D_{i'j'} = \varnothing$ for $(i,j) \neq (i',j')$.

Finally, an $(n, M, N, \lambda)$ code for the MAC, $W$, is an $(n, M, N)$ code with

$$\frac{1}{MN} \sum_{i=1}^{M} \sum_{j=1}^{N} W^n(D_{i,j}|\mathbf{u}_i, \mathbf{v}_j) \geq 1 - \lambda. \tag{10}$$

## 3 main result

The main result of this paper is a lower (*sphere packing*) bound for the maximal error probability for a MAC. To state the new bound we need an intermediate result that has the form of a strong converse for the MAC. We state this result here and relegate the proof to the appendix.

**Definition 1.** *For any DM-MAC, $W$, for any joint distribution $P \in \mathcal{P}(\mathcal{X} \times \mathcal{Y})$, any $0 \leq \lambda < 1$, and any $(n, M, N)$ code, $C$, define*

$$\mathcal{E}_W(P, \lambda) \triangleq \{(\mathbf{u}_i, \mathbf{v}_j) \in C : W(D_{ij}|\mathbf{u}_i, \mathbf{v}_j) \geq \frac{1-\lambda}{2},$$
$$(\mathbf{u}_i, \mathbf{v}_j) \in T_P\}. \tag{11}$$

**Theorem 1.** *Consider any $(n, M, N)$ code $C$. For every $P_{XY}^n \in \mathcal{P}_n(\mathcal{X} \times \mathcal{Y})$, such that $|\mathcal{E}_W(P_{XY}^n, \lambda)| \geq \frac{1}{(n+1)^{|\mathcal{X}||\mathcal{Y}|}}(1 - \frac{2\lambda}{1+\lambda})MN$, then*

$$(\frac{1}{n}\log M, \frac{1}{n}\log N) \in C_W^n(P_{XY}^n) \tag{12}$$

where $C_W^n(P)$ is defined as the closure of the set of all $(R_1, R_2)$ pairs satisfying

$$R_1 \leq I(\bar{X} \wedge \bar{Z}|\bar{Y}, Q) + \epsilon_n \tag{13a}$$
$$R_2 \leq I(\bar{Y} \wedge \bar{Z}|\bar{X}, Q) + \epsilon_n \tag{13b}$$
$$R_1 + R_2 \leq I(\bar{X}\bar{Y} \wedge \bar{Z}|Q) + \epsilon_n \tag{13c}$$

for some choice of random variables $Q$ defined on $\{1, 2, 3, 4\}$, and joint distribution $p(q)p(x|q)p(y|q)w(z|x,y)$, with marginal distribution $p(x,y) = P^n(x,y)$. Here, $\epsilon_n \to 0$ an $n \to \infty$.

We further define $C_W(P)$ (the *limiting* version of the sets $C_W^n(P)$) as the closure of the set of all $(R_1, R_2)$ pairs satisfying

$$R_1 \leq I(\bar{X} \wedge \bar{Z}|\bar{Y}, Q) \tag{14a}$$
$$R_2 \leq I(\bar{Y} \wedge \bar{Z}|\bar{X}, Q) \tag{14b}$$
$$R_1 + R_2 \leq I(\bar{X}\bar{Y} \wedge \bar{Z}|Q), \tag{14c}$$

for some choice of random variables $Q$ defined on $\{1, 2, 3, 4\}$, and joint distribution $p(q)p(x|q)p(y|q)w(z|x,y)$, with marginal distribution $p(x,y) = P(x,y)$.

**Theorem 2.** *(Sphere Packing Bound). For any $R_X, R_Y > 0$, $\delta > 0$ and any DM-MAC, $W : \mathcal{X} \times \mathcal{Y} \to \mathcal{Z}$, every $(n, M, N, \lambda)$ code, $C$ with*

$$\frac{1}{n}\log M \geq R_X + \delta \tag{15a}$$
$$\frac{1}{n}\log N \geq R_Y + \delta, \tag{15b}$$

*has maximum probability of error*

$$P_e^m \geq \frac{1}{2}\exp\big(-nE_{sp}(R_X, R_Y, W)(1+\delta)\big), \tag{16}$$

*where*

$$E_{sp}(R_X, R_Y, W) \triangleq \max_{P_{XY} \in \mathcal{P}(\mathcal{X} \times \mathcal{Y})} \min_{\substack{V:(R_X, R_Y) \\ \notin C_V(P_{XY})}} D(V||W|P_{XY}). \tag{17}$$

*Proof.* If $\lambda = 1$, the result is trivial. Assume $\lambda < 1$. Let's choose $\lambda'$ such that $\max\{1-\delta, \lambda\} < \lambda' < 1$. Since $\lambda' > \lambda$, every $(n, M, N, \lambda)$ code is also an $(n, M, N, \lambda')$ code. Using the same argument as [4, pp. 189], we conclude that there exist at least one dominant type $P^n \in \mathcal{P}_n(\mathcal{X} \times \mathcal{Y})$, such that $|\mathcal{E}_W(P^n, \lambda')| \geq \frac{1}{(n+1)^{|\mathcal{X}||\mathcal{Y}|}}(1 - \frac{2\lambda'}{1+\lambda'})MN$. Consider an arbitrary DM-MAC $V : \mathcal{X} \times \mathcal{Y} \to \mathcal{Z}$, such that $(R_X, R_Y) \notin C_V^n(P^n)$. By Theorem 1, there exist at least one pair $(\mathbf{u}_i, \mathbf{v}_j)$ with joint type $P_{XY}^n$ such that

$$V^n(D_{ij}^c | \mathbf{u}_i, \mathbf{v}_j) > \frac{1 + \lambda'}{2} > 1 - \frac{\delta}{2}. \tag{18}$$

Using the same method as Csiszar in [5, pp. 167], we have

$$W^n(D_{ij}^c | \mathbf{u}_i, \mathbf{v}_j) \geq \exp\left\{-\frac{D(V||W|P^n) + h(1 - \frac{\delta}{2})}{1 - \frac{\delta}{2}}\right\}$$

$$\geq \frac{1}{2} \exp\{-nD(V||W|P^n)(1 + \delta)\}, \tag{19}$$

for small enough $\delta$ satisfying $h(1 - \frac{\delta}{2}) < 1 - \frac{\delta}{2}$. By maximizing the result over the arbitrary channel $V$, we get

$$P_e^m \geq \max_{\substack{V:(R_X, R_Y) \\ \notin C_V^n(P^n)}} \frac{1}{2} \exp\{-nD(V||W|P^n)(1+\delta)\}$$

$$= \frac{1}{2} \exp\{-n \min_{\substack{V:(R_X, R_Y) \\ \notin C_V^n(P^n)}} D(V||W|P^n)(1+\delta)\}$$

$$\geq \min_{\substack{P^n \in \\ \mathcal{P}_n(\mathcal{X} \times \mathcal{Y})}} \frac{1}{2} \exp\{-n \min_{\substack{V:(R_X, R_Y) \\ \notin C_V^n(P^n)}} D(V||W|P^n)(1+\delta)\}$$

$$\geq \min_{\substack{P \in \\ \mathcal{P}(\mathcal{X} \times \mathcal{Y})}} \frac{1}{2} \exp\{-n \min_{\substack{V:(R_X, R_Y) \\ \notin C_V^n(P)}} D(V||W|P)(1+\delta)\} \tag{20}$$

Using Lemma 6, we conclude that for sufficiently large $n$,

$$P_e^m \geq$$

$$\min_{\substack{P \in \\ \mathcal{P}(\mathcal{X} \times \mathcal{Y})}} \frac{1}{2} \exp\{-n \min_{\substack{V:(R_X, R_Y) \\ \notin C_V(P)}} D(V||W|P)(1+\delta)\}. \tag{21}$$

□

which completes the proof.

## 4 appendix

The basic idea of the proof is wringing technique which was used for the first time, by Ahlswede [6].

Consider any $P_{XY}^n \in \mathcal{P}_n(\mathcal{X} \times \mathcal{Y})$, such that $|\mathcal{E}_W(P_{XY}^n, \lambda)| \geq \frac{1}{(n+1)^{|\mathcal{X}||\mathcal{Y}|}}(1 - \frac{2\lambda}{1+\lambda})MN$. let's define $\mathcal{A} \triangleq \{(i,j) : W(D_{ij}|\mathbf{u}_i, \mathbf{v}_j) \geq \frac{1-\lambda}{2}, (\mathbf{u}_i, \mathbf{v}_j) \in T_{P_{XY}^n}\}$. Since $|\mathcal{A}| = |\mathcal{E}_W(P_{XY}^n, \lambda)|$, we conclude that

$$|\mathcal{A}| \geq \frac{1}{(n+1)^{|\mathcal{X}||\mathcal{Y}|}}(1 - \frac{2\lambda}{1+\lambda})MN. \tag{22}$$

Define,

$$\mathcal{C}(i) = \{(i,j) : (i,j) \in \mathcal{A}, 1 \leq j \leq N\} \tag{23a}$$
$$\mathcal{B}(j) = \{(i,j) : (i,j) \in \mathcal{A}, 1 \leq i \leq M\}. \tag{23b}$$

Consider the subcode $\{(\mathbf{u}_i, \mathbf{v}_j, D_{ij}) : (i,j) \in \mathcal{A}\}$ and define random variables $X^n, Y^n$

$$Pr((X^n, Y^n) = (\mathbf{u}_i, \mathbf{v}_j)) = \frac{1}{|\mathcal{A}|} \text{if } (i,j) \in \mathcal{A}. \tag{24}$$

**Lemma 1.** *For random variables $X^n$, $Y^n$ defined in (24), the mutual information satisfies the following inequality:*

$$I(X^n \wedge Y^n) \leq -\log(1 - \frac{2\lambda}{1+\lambda}) + |\mathcal{X}||\mathcal{Y}|\log(n+1). \tag{25}$$

*Proof.* This is a generalization of the proof by Dueck in [4]. Observe that

$$H(Y^n|X^n) = \sum_{\mathbf{u}_i} Pr(X^n = \mathbf{u}_i) H(Y^n|\mathbf{u}_i). \tag{26}$$

However, by the definition of the variables $X^n$, $Y^n$ we have

$$H(Y^n|\mathbf{u}_i) = \log |\{j : (i,j) \in \mathcal{A}\}| \tag{27}$$

and

$$Pr(X^n = \mathbf{u}_i) = |\mathcal{A}|^{-1} \cdot |\{j : (i,j) \in \mathcal{A}\}|. \tag{28}$$

Hence,

$$H(Y^n|X^n) = |\mathcal{A}|^{-1} \sum_{i=1}^{M} |\{j : (i,j) \in \mathcal{A}\}| \log |\{j : (i,j) \in \mathcal{A}\}|. \tag{29}$$

In the right hand side of (29), the summands are of the form $m \log m$. This function of $m$ is increasing and convex in $m$. Thus,

$$H(Y^n|X^n) \geq$$
$$|\mathcal{A}|^{-1}(\sum_{i=1}^{M} |\{j : (i,j) \in \mathcal{A}\}|) \log(M^{-1} \sum_{i=1}^{M} |\{j : (i,j) \in \mathcal{A}\}|), \tag{30}$$

and since

$$\sum_{i=1}^{M}|\{j:(i,j)\in\mathcal{A}\}|=|\mathcal{A}|, \tag{31}$$

we have

$$H(Y^n|X^n)\geq \log(M^{-1}|\mathcal{A}|). \tag{32}$$

By (22), we conclude that

$$H(Y^n|X^n)\geq$$
$$\log N+\log(1-\frac{2\lambda}{1+\lambda})-|\mathcal{X}||\mathcal{Y}|\log(n+1). \tag{33}$$

Finally,

$$I(X^n\wedge Y^n)=H(Y^n)-H(Y^n|X^n)$$
$$\leq \log N-H(Y^n|X^n), \tag{34}$$

which concludes the proof. □

**Lemma 2.** *[7] Let $X^n$, $Y^n$ be RV's with values in $\mathcal{X}^n$, $\mathcal{Y}^n$ resp. and assume that*

$$I(X^n\wedge Y^n)\leq \sigma \tag{35}$$

*Then, for any $0<\delta<\sigma$ there exist $t_1,t_2,...,t_k\in\{1,...,n\}$ where $0\leq k<\frac{2\sigma}{\delta}$ such that for some $\bar{x}_{t_1},\bar{y}_{t_1},\bar{x}_{t_2},\bar{y}_{t_2},...,\bar{x}_{t_k},\bar{y}_{t_k}$*

$$I(X_t\wedge Y_t|X_{t_1}=\bar{x}_{t_1},Y_{t_1}=\bar{y}_{t_1},...,X_{t_k}=\bar{x}_{t_k},Y_{t_k}=\bar{y}_{t_k})\leq \delta$$
$$\text{for } t=1,2,...,n \tag{36}$$

*and*

$$Pr(X_{t_1}=\bar{x}_{t_1},Y_{t_1}=\bar{y}_{t_1},...,X_{t_k}=\bar{x}_{t_k},Y_{t_k}=\bar{y}_{t_k})$$
$$\geq (\frac{\delta}{|\mathcal{X}||\mathcal{Y}|(2\sigma-\delta)})^k. \tag{37}$$

Consider the subcode $\{(\mathbf{u}_i,\mathbf{v}_j,D_{ij}):(i,j)\in\bar{\mathcal{A}}\}$, where

$$\bar{\mathcal{A}}\triangleq \{(i,j)\in\mathcal{A}:\mathbf{u}_{it_l}=\bar{x}_{t_l},\mathbf{v}_{jt_l}=\bar{y}_{t_l}\ 1\leq l\leq k\} \tag{38}$$

and define

$$\bar{\mathcal{C}}(i)=\{(i,j):(i,j)\in\bar{\mathcal{A}},1\leq j\leq N\} \tag{39a}$$
$$\bar{\mathcal{B}}(j)=\{(i,j):(i,j)\in\bar{\mathcal{A}},1\leq i\leq M\}. \tag{39b}$$

**Lemma 3.** *The subcode* $\{(\mathbf{u}_i, \mathbf{v}_j, D_{ij}) : (i,j) \in \bar{\mathcal{A}}\}$, *is a subcode with maximal error probability* $\frac{1+\lambda}{2}$, *and*

$$|\bar{\mathcal{A}}| \geq (\frac{\delta}{|\mathcal{X}||\mathcal{Y}|(2\sigma - \delta)})^k |\mathcal{A}|. \tag{40}$$

*Moreover,*

$$\sum_{x,y} |Pr(X_t = x, Y_t = y) - Pr(X_t = x)Pr(Y_t = y)| \leq 2\delta^{1/2}, \tag{41}$$

*where* $X^n = (X_1, ..., X_n)$, $Y^n = (Y_1, ..., Y_n)$ *are distributed according to the Fano-distribution of the subcode* $\{(\mathbf{u}_i, \mathbf{v}_j, D_{ij}) : (i,j) \in \bar{\mathcal{A}}\}$.

*Proof.* Since $\bar{\mathcal{A}} \subset \mathcal{A}$, the maximal probability of error for this subcode is at most $\frac{1+\lambda}{2}$. The second part of Lemma 2, yields immediately (40). On the other hand,

$$\begin{aligned}
&P_{\mathcal{A}}(X_t = x, Y_t = y | \bar{x}_{t_1}, \bar{y}_{t_1}, \bar{x}_{t_2}, \bar{y}_{t_2}, ..., \bar{x}_{t_k}, \bar{y}_{t_k}) \\
&= \frac{P_{\mathcal{A}}(X_t = x, Y_t = y, \bar{x}_{t_1}, \bar{y}_{t_1}, \bar{x}_{t_2}, \bar{y}_{t_2}, ..., \bar{x}_{t_k}, \bar{y}_{t_k})}{P_{\mathcal{A}}(\bar{x}_{t_1}, \bar{y}_{t_1}, \bar{x}_{t_2}, \bar{y}_{t_2}, ..., \bar{x}_{t_k}, \bar{y}_{t_k})} \\
&= \frac{N_{\mathcal{A}}(X_t = x, Y_t = y, \bar{x}_{t_1}, \bar{y}_{t_1}, \bar{x}_{t_2}, \bar{y}_{t_2}, ..., \bar{x}_{t_k}, \bar{y}_{t_k})}{N_{\mathcal{A}}(\bar{x}_{t_1}, \bar{y}_{t_1}, \bar{x}_{t_2}, \bar{y}_{t_2}, ..., \bar{x}_{t_k}, \bar{y}_{t_k})} \\
&= \frac{N_{\bar{\mathcal{A}}}(X_t = x, Y_t = y)}{|\bar{\mathcal{A}}|} \\
&= P_{\bar{\mathcal{A}}}(X_t = x, Y_t = y). \tag{42}
\end{aligned}$$

Therefore, by the first part of Lemma 2, we conclude that

$$I(X_t \wedge Y_t) \leq \delta, \quad \text{for } 1 \leq t \leq n. \tag{43}$$

Since $I(X_t \wedge Y_t)$ is an *I-divergence*, Pinsker's inequality implies [8]

$$\sum_{x,y} |Pr(X_t = x, Y_t = y) - Pr(X_t = x)Pr(Y_t = y)| \leq 2\delta^{1/2}. \tag{44}$$

□

**Lemma 4.** *[9]: For a* $(n, M, \lambda)$ *code* $\{(\mathbf{u}_i, D_i) : 1 \leq i \leq M\}$ *for the non-stationary DMC* $(W_t)_{t=1}^{\infty}$

$$\log M < \sum_{t=1}^{n} I(X_t \wedge Z_t) + \frac{3}{1-\lambda} |\mathcal{X}| n^{1/2}, \tag{45}$$

*where the distribution of the RV's are determined by the Fano-distribution on the codewords.*

Define random variables $\bar{X}^n, \bar{Y}^n$ on $\mathcal{X}^n$ resp. $\mathcal{Y}^n$ by

$$Pr((\bar{X}^n, \bar{Y}^n) = (\mathbf{u}_i, \mathbf{v}_j)) = \frac{1}{|\bar{\mathcal{A}}|} \text{if } (i,j) \in \bar{\mathcal{A}}. \tag{46}$$

**Lemma 5.** *For any $0 \leq \lambda < 1$, any $(n, M, N)$ code $C \triangleq \{(\mathbf{u}_i, \mathbf{v}_j, D_{ij}) : 1 \leq i \leq M, 1 \leq j \leq N\}$ for the any MAC, $W$, and for any $P_{XY}^n \in \mathcal{P}_n(\mathcal{X} \times \mathcal{Y})$, such that $|\mathcal{E}_W(P_{XY}^n, \lambda)| \geq \frac{1}{(n+1)^{|\mathcal{X}||\mathcal{Y}|}}(1 - \frac{2\lambda}{1+\lambda})MN$*

$$\log M \leq \sum_{t=1}^{n} I(\bar{X}_t \wedge \bar{Z}_t | \bar{Y}_t) + c_1(\lambda) n^{1/2} + c_1 k \log(\frac{2\sigma}{\delta})$$

$$\log N \leq \sum_{t=1}^{n} I(\bar{Y}_t \wedge \bar{Z}_t | \bar{X}_t) + c_2(\lambda) n^{1/2} + c_2 k \log(\frac{2\sigma}{\delta})$$

$$\log MN \leq \sum_{t=1}^{n} I(\bar{X}_t \bar{Y}_t \wedge \bar{Z}_t) + c_3(\lambda) n^{1/2} + c_3 k \log(\frac{2\sigma}{\delta}),$$

*where the distributions of the RV's are determined by the Fano-distribution on the codewords $\{(\mathbf{u}_i, \mathbf{v}_j) : (i,j) \in \bar{\mathcal{A}}\}$. Here, $c_i(\lambda)$ and $c_i$ are suitable functions of $\lambda$.*

*Proof.* For any fixed $j$, consider $(n, |\bar{\mathcal{B}}(j)|)$ code $\{(\mathbf{u}_i, D_{ij}) : (i,j) \in \bar{\mathcal{B}}(j)\}$. Any pair of codewords in this code has probability of error at most equal to $\frac{1+\lambda}{2}$. Let's define $\lambda' \triangleq \frac{1+\lambda}{2}$. It follows from Lemma 4 that

$$\log |\bar{\mathcal{B}}(j)| \leq \sum_{t=1}^{n} I(\bar{X}_t \wedge \bar{Z}_t | \bar{Y}_t = \mathbf{v}_{jt}) + \frac{3}{1-\lambda'} |\mathcal{X}| n^{1/2}. \tag{47}$$

Similarly,

$$\log |\bar{\mathcal{C}}(i)| \leq \sum_{t=1}^{n} I(\bar{Y}_t \wedge \bar{Z}_t | \bar{X}_t = \mathbf{u}_{it}) + \frac{3}{1-\lambda'} |\mathcal{Y}| n^{1/2} \tag{48}$$

$$\log |\bar{\mathcal{A}}| \leq \sum_{t=1}^{n} I(\bar{X}_t \bar{Y}_t \wedge \bar{Z}_t) + \frac{3}{1-\lambda'} |\mathcal{X}||\mathcal{Y}| n^{1/2}. \tag{49}$$

Since $Pr(\bar{Y}_t = y) = |\bar{\mathcal{A}}|^{-1} \sum_{(i,j) \in \bar{\mathcal{A}}} 1_{\{\mathbf{v}_{jt}, y\}}$,

$$|\bar{\mathcal{A}}|^{-1} \sum_{(i,j) \in \bar{\mathcal{A}}} \log |\bar{\mathcal{B}}(j)|$$
$$\leq \sum_{(i,j) \in \bar{\mathcal{A}}} \sum_{t=1}^{n} I(\bar{X}_t \wedge \bar{Z}_t | \bar{Y}_t = \mathbf{v}_{jt}) \frac{\sum_y 1_{\{\mathbf{v}_{jt}, y\}}}{|\bar{\mathcal{A}}|}$$
$$+ \frac{3}{1-\lambda'} |\mathcal{X}| n^{1/2}$$
$$= \sum_{t=1}^{n} I(\bar{X}_t \wedge \bar{Z}_t | \bar{Y}_t) + \frac{3}{1-\lambda'} |\mathcal{X}| n^{1/2}. \tag{50}$$

Define $\lambda^* \triangleq \frac{2\lambda}{1+\lambda}$, and

$$B^* \triangleq \frac{1-\lambda^*}{n} \frac{M}{(n+1)^{|\mathcal{X}||\mathcal{Y}|}} \left(\frac{\delta}{|\mathcal{X}||\mathcal{Y}|(2\sigma-\delta)}\right)^k. \quad (51)$$

Therefore,

$$\begin{aligned}
|\bar{\mathcal{A}}|^{-1} &\sum_{(i,j)\in\bar{\mathcal{A}}} \log|\bar{\mathcal{B}}(j)| \\
&= |\bar{\mathcal{A}}|^{-1} \sum_j |\bar{\mathcal{B}}(j)| \log|\bar{\mathcal{B}}(j)| \\
&\geq |\bar{\mathcal{A}}|^{-1} \sum_{j:|\bar{\mathcal{B}}(j)|\geq B^*} |\bar{\mathcal{B}}(j)| \log|\bar{\mathcal{B}}(j)| \\
&\geq |\bar{\mathcal{A}}|^{-1} \log(B^*) \sum_{j:|\bar{\mathcal{B}}(j)|\geq B^*} |\bar{\mathcal{B}}(j)| \\
&\geq |\bar{\mathcal{A}}|^{-1} \log(B^*)(|\bar{\mathcal{A}}| - NB^*). \quad (52)
\end{aligned}$$

By lemma 3, (22), and definition of $B^*$,

$$NB^* \leq \frac{1}{n}|\bar{\mathcal{A}}|. \quad (53)$$

Therefore,

$$\begin{aligned}
|\bar{\mathcal{A}}|^{-1} &\sum_{(i,j)\in\bar{\mathcal{A}}} \log|\bar{\mathcal{B}}(j)| \\
&\geq |\bar{\mathcal{A}}|^{-1} \log(B^*)(|\bar{\mathcal{A}}| - \frac{1}{n}|\bar{\mathcal{A}}|) \\
&= (1 - \frac{1}{n}) \log\left(\frac{1-\lambda^*}{n} \frac{M}{(n+1)^{|\mathcal{X}||\mathcal{Y}|}} \left(\frac{\delta}{|\mathcal{X}||\mathcal{Y}|(2\sigma-\delta)}\right)^k\right). \quad (54)
\end{aligned}$$

By (50), (54)

$$\begin{aligned}
\log M &\leq (1+\frac{2}{n})\left(\sum_{t=1}^n I(\bar{X}_t \wedge \bar{Z}_t|\bar{Y}_t) + \frac{3}{1-\lambda'}|\mathcal{X}|n^{1/2}\right) \\
&\quad - \log(1-\lambda^*) + \log n + |\mathcal{X}||\mathcal{Y}|\log(n+1) \\
&\quad + k\log\left(\frac{|\mathcal{X}||\mathcal{Y}|2\sigma}{\delta}\right) \\
&\leq \sum_{t=1}^n I(\bar{X}_t \wedge \bar{Z}_t|\bar{Y}_t) + c_1(\lambda')n^{1/2} + c_1 k\log\left(\frac{2\sigma}{\delta}\right) \\
&\quad + 2|\mathcal{Z}| \quad (55)
\end{aligned}$$

Analogously,

$$\begin{aligned}
\log N &\leq \sum_{t=1}^n I(\bar{Y}_t \wedge \bar{Z}_t|\bar{X}_t) + c_2(\lambda')n^{1/2} + c_2 k\log\left(\frac{2\sigma}{\delta}\right) \\
&\quad + 2|\mathcal{Z}|. \quad (56)
\end{aligned}$$

To find an upper bound for $\log MN$, we first try to find a lower bound on the $\log |\bar{\mathcal{A}}|$. By Lemma 3

$$\log |\bar{\mathcal{A}}| \geq \log |\mathcal{A}| + k \log(\frac{\delta}{|\mathcal{X}||\mathcal{Y}|(2\sigma - \delta)})$$

$$\geq \log |\mathcal{A}| + k \log(\frac{\delta}{|\mathcal{X}||\mathcal{Y}|2\sigma})$$

$$= \log |\mathcal{A}| - k \log(\frac{2\sigma}{\delta}) - k \log(|\mathcal{X}||\mathcal{Y}|)$$

$$\geq \log(MN) - |\mathcal{X}||\mathcal{Y}| \log(n+1) + \log(1 - \frac{2\lambda}{1+\lambda})$$

$$- k \log(\frac{2\sigma}{\delta}) - k \log(|\mathcal{X}||\mathcal{Y}|). \tag{57}$$

Therefore,

$$\log(MN) \leq \log |\bar{\mathcal{A}}| + c_3 k \log(\frac{2\sigma}{\delta}). \tag{58}$$

Using (49),

$$\log MN \leq \sum_{t=1}^{n} I(\bar{X}_t \bar{Y}_t \wedge \bar{Z}_t) + c_3(\lambda') n^{1/2} + c_3 k \log(\frac{2\sigma}{\delta}). \tag{59}$$

$\square$

Note that, in general $\bar{X}_t$ and $\bar{Y}_t$ are not independent. In the following, we prove that they are nearly independent.

Now, we combine (25) and lemma 3. For an $(n, M, N)$ code $\{(\mathbf{u}_i, \mathbf{v}_j, D_{ij}) : 1 \leq i \leq M, 1 \leq j \leq N\}$ which has the particular property mentioned in theorem 1, define $\mathcal{A}, \bar{\mathcal{A}}$ as defined before. Apply lemma 3 with parameter $\delta = n^{-1/2}$. Using $\sigma = -\log(1 - \frac{2\lambda}{1+\lambda}) + |\mathcal{X}||\mathcal{Y}| \log(n+1)$, we conclude that

$$k \leq \frac{2\sigma}{\delta} = 2\sqrt{n}(-\log(1 - \frac{2\lambda}{1+\lambda}) + |\mathcal{X}||\mathcal{Y}| \log(n+1))$$

$$\sim O(\sqrt{n} \log n) \tag{60}$$

and

$$|Pr(\bar{X}_t = x, \bar{Y}_t = y) - Pr(\bar{X}_t = x) Pr(\bar{Y}_t = y)| \leq 2n^{-1/4}, \tag{61}$$

for any $x \in \mathcal{X}$, $y \in \mathcal{Y}$, and $t = 1, ..., n$. We can write the above equations as follows

$$\frac{1}{n} \log M \leq \frac{1}{n} \sum_{t=1}^{n} I(\bar{X}_t \wedge \bar{Z}_t | \bar{Y}_t) + C(\lambda) \frac{o(n)}{n} \tag{62a}$$

$$\frac{1}{n} \log N \leq \frac{1}{n} \sum_{t=1}^{n} I(\bar{Y}_t \wedge \bar{Z}_t | \bar{X}_t) + C(\lambda) \frac{o(n)}{n} \tag{62b}$$

$$\frac{1}{n} \log MN \leq \frac{1}{n} \sum_{t=1}^{n} I(\bar{X}_t \bar{Y}_t \wedge \bar{Z}_t) + C(\lambda) \frac{o(n)}{n}. \tag{62c}$$

The expressions in (62a)-(62c) are the averages of the mutual informations calculated at the empirical distributions in the column $t$ of the mentioned subcode. We can rewrite these equations with the new variable $Q$, where $Q = q \in \{1, 2, ..., n\}$ with probability $\frac{1}{n}$. Using the same method as Cover [1, pp. 402], we obtain the result. The only thing remained to be found is the distribution under which we calculate the mutual informations. However, by (61)

$$\begin{aligned}
|P(\bar{X} = x, \bar{Y} = y|Q = q) - P(\bar{X} = x|Q = q)P(\bar{Y} = y|Q = q)| \\
= |Pr(\bar{X}_q = x, \bar{Y}_q = y) - Pr(\bar{X}_q = x)Pr(\bar{Y}_q = y)| \\
\leq 2n^{-1/4}.
\end{aligned} \quad (63)$$

Using the continuity of conditional mutual information with respect to distributions, using the same idea of [10, pp. 722], we conclude that, if two distributions are close, the conditional mutual informations, calculated based on them, cannot be too far. More precisely, we can say that there exists a sequence $\{\delta_n\}_{n=1}^{\infty}$, $\delta_n \to 0$ as $n \to \infty$, such that,

$$\begin{aligned}
\frac{1}{n} \log M &\leq I(\bar{X}_t \wedge \bar{Z}_t | \bar{Y}_t, Q) + C(\lambda)\frac{o(n)}{n} + \delta_n \\
\frac{1}{n} \log N &\leq I(\bar{Y}_t \wedge \bar{Z}_t | \bar{X}_t, Q) + C(\lambda)\frac{o(n)}{n} + \delta_n \\
\frac{1}{n} \log MN &\leq I(\bar{X}_t \bar{Y}_t \wedge \bar{Z}_t, Q) + C(\lambda)\frac{o(n)}{n} + \delta_n.
\end{aligned} \quad (64)$$

Here, the mutual informations calculated based on $p(q)p(x|q)p(y|q)w(z|x,y)$, with marginal distribution $P_{XY}^n(x,y)$. On the other hand, the joint probability distribution of $\bar{X}$ and $\bar{Y}$ is

$$\begin{aligned}
P(\bar{X} = x, \bar{Y} = y) \\
= \sum_{(i,j) \in \bar{\mathcal{A}}} P(\bar{X}(W_1) = x, \bar{Y}(W_2) = y, W_1 = i, W_2 = j) \\
= \sum_{(i,j) \in \bar{\mathcal{A}}} P(\bar{X}(i) = x, \bar{Y}(j) = y)P(i,j) \\
= \frac{1}{|\bar{\mathcal{A}}|} \sum_{(i,j) \in \bar{\mathcal{A}}} P(\bar{X}(i) = x, \bar{Y}(j) = y) \\
= \frac{1}{|\bar{\mathcal{A}}|} \sum_{(i,j) \in \bar{\mathcal{A}}} \frac{1}{n} \sum_{q=1}^{n} 1\{\bar{X}_q(i) = x, \bar{Y}_q(j) = y\}
\end{aligned} \quad (65)$$

However, all codewords have the same joint type $P_{XY}^n$, therefore,

$$\sum_{q=1}^{n} 1\{\bar{X}_q(i) = x, \bar{Y}_q(j) = y\} = nP_{XY}^n(x, y). \quad (66)$$

(65) and (66) result in

$$P(\bar{X} = x, \bar{Y} = y) = P_{XY}^n(x, y). \tag{67}$$

Finally, we can conclude that

$$P(q, x, y, z) = p(q)p(x|q)p(y|q)W(z|x, y), \tag{68}$$

in which the marginal distribution of $\bar{X}$ and $\bar{Y}$ is $P_{XY}^n(x, y)$.

The cardinality bound on the time-sharing random variable, $Q$, is the consequence of Carathéodory's theorem on the convex set [11], [12], [1].

**Lemma 6.** *For any fixed $P \in \mathcal{P}(\mathcal{X} \times \mathcal{Y})$, rate pair $(R_X, R_Y)$,*

$$\lim_{n \to \infty} \min_{V \in D^n(P)} D(V||W|P) = \min_{V \in D(P)} D(V||W|P), \tag{69}$$

*where*

$$\begin{aligned} D(P) &\triangleq \{V : (R_X, R_Y) \notin C_V(P)\} \\ D^n(P) &\triangleq \{V : (R_X, R_Y) \notin C_V^n(P)\}. \end{aligned} \tag{70}$$

*Proof.* Define $\alpha_n \triangleq \min_{V \in D^n(P)} D(V||W|P)$, and $\alpha^* \triangleq \min_{V \in D(P)} D(V||W|P)$. Moreover, suppose $\alpha^*$ is achieved by $V^*$. Since $\{\alpha_n\}_{n=1}^\infty$ is a decreasing sequence and it is bounded from below ($\alpha_n \geq \alpha^*$), therefore it has a limit. Suppose the limit is not equal to $\alpha^*$. Therefore, there exist a $\delta > 0$, such that for all sufficiently large $n$,

$$|\alpha_n - \alpha^*| \geq \delta. \tag{71}$$

Hence, for all $V \in D^n(P)$, for all sufficiently large $n$,

$$D(V||W|P) - \alpha^* \geq \delta \tag{72}$$

which concludes that $V^*$ cannot belong to $D^n(P)$, i.e., for all sufficiently large $n$,

$$(R_X, R_Y) \in C_{V^*}^n(P). \tag{73}$$

Since $V^* \in D(P)$,

$$(R_X, R_Y) \notin C_{V^*}(P). \tag{74}$$

Therefore $C_{V^*}^n(P)$ cannot converge to $C_{V^*}(P)$, which is a contradiction. □

# References

[1] T. M. Cover and J. A. Thomas, *Elements of Information Theory*. New York: John Wiley & Sons, 1991.

[2] D. Slepian and J. K. Wolf, "A coding theorem for multiple access channels with correlated sources," *bell Syst. tech. J.*, vol. 52, pp. 1037–1076, 1973.

[3] E. A. Haroutunian, "Lower bound for the error probability of multiple-access channels," *Problemy Peredachi Informatsii*, vol. 11, pp. 23–36, June 1975.

[4] G. Dueck, "The strong converse of the coding theorem for the multi-user channels," *Journal of Combinatorics, Information and System sciences*, pp. 187–196, 1981.

[5] I. Csiszar and J. Korner, *Information theory: Coding theorems for Discrete memoryless Systems.*, 1981.

[6] R. Ahlswede, "On two-way communication channels and a problem by zarankiewics," in *Probl. of Control and Inform. Theory.*

[7] ——, "An elementary proof of the strong converse theorem for the multiple access channel," *Journal of Combinatorics, Information and System sciences*, pp. 216–230, 1982.

[8] P. H. A. Fedotov and F. Topsøe, "Refinements of pinsker's inequality," *IEEE Trans. Information Theory*, vol. 49, no. 6, pp. 1491–1498, June 2003.

[9] U. Augustin, "Gedachtnisfreie kannale for diskrete zeit," *Z. Wahrscheinlichkelts theory verw*, pp. 10–61, 1966.

[10] R. Ahlswede, "The rate-distortion for multiple description without excess rates," *IEEE Trans. Information Theory*, vol. 31, no. 6, pp. 721–726, Nov. 1985.

[11] H. G. Eggleston, *Convexity.* Cambridge, UK: Cambridge University Press, 1969.

[12] B. Grünbaum, *Convex Polytopes.* New York: McGraw-Hill, 1976.